\documentstyle[11pt,newpasp,epsf]{article}

\begin{document}

\title{Constraints on Galaxy Formation from the Tully-Fisher Relation}
\author{Frank C. van den Bosch}
\affil{Department of Astronomy, University of Washington, Seattle}

\begin{abstract}
  New models for   the formation of  disk  galaxies  are presented.  I
  discuss the constraints on galaxy  formation that follow from fitting
  the  model to the near-infrared Tully-Fisher  (TF) relation, with an
  emphasis on reproducing the  small amount of scatter  observed. Once
  the parameters that describe the supernova feedback are tuned to fit
  the  slope of  the observed TF  relation,  the model reproduces  the
  correct    amount of TF   scatter, and  yields   gas mass fractions,
  mass-to-light ratios, and  characteristic accelerations that are all
  in excellent agreement with observations. 
\end{abstract}


\section{Introduction}

Understanding  the  formation  of  galaxies  is intimately  related to
understanding the  origin  of the  fundamental scaling  relations.  In
particular, any  successful theory for  the formation  of disk galaxies
should be able to  explain the slope, zero-point  and small  amount of
scatter of the Tully-Fisher relation  (TFR).  The empirical TFR  which
most directly reflects the mass in stars  and the total dynamical mass
of the halo  is the $K$-band TFR  of Verheijen (1997), which  uses the
flat part of  the rotation curve as velocity  measure, and we use this
relation to constrain our models.

\section{Modeling the Formation of Disk Galaxies}

We  assume  disks  to form by    the settling  of  baryonic matter  in
virialized dark  halos described by the  NFW density profile (Navarro,
Frenk \& White 1997).   It   is assumed  that baryons conserve   their
angular momentum, thus settling  into a  disk  (cf.  Mo, Mao \&  White
1998).  Adiabatic contraction of the  dark halo is taken into account,
as well  as a  recipe for bulge  formation based  on a self-regulating
mechanism that ensures disks to be stable (van  den Bosch 1998).  Once
the density distribution of the baryonic material is known, we compute
the fraction of baryons converted into stars.  Only gas with densities
above  the critical density given by  Toomre's  stability criterion is
considered eligible for star formation (cf.  Kennicutt 1989). A simple
recipe  for supernovae   feedback is included,    which describes what
fraction of  the baryonic mass is  prevented from becoming part of the
disk/bulge system. The slope and scatter of the TFR depend strongly on
the luminosity and velocity measures used.  Therefore, it is essential
that one extracts the same measures from the models as the ones in the
TFR used to constrain those  models.  We improve upon previous studies
by carefully doing so.  Details of the models can  be found in van den
Bosch (1999) and van den Bosch \& Dalcanton (1999).

\begin{figure}
\plotone{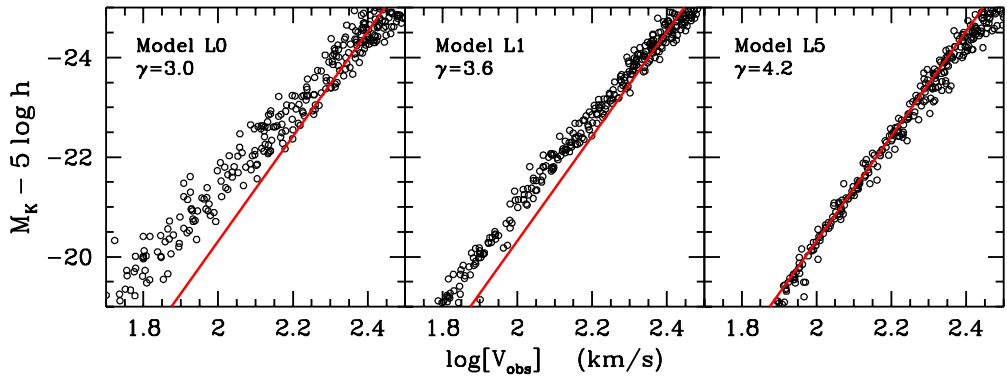}
\caption{TFRs for   three  models  (open  circles)  compared   to  the
empirical $K$-band TFR of Verheijen (1997, solid line). See the text
for details.}
\end{figure}

\section{Results}

Within the framework of dark matter, simple  dynamics predict a TFR of
the form $L  \propto V_{\rm  rot}^{\gamma}$ with  $\gamma =  3$.   The
empirical  $K$-band TFR,  however,   has $\gamma \simeq  4.2$.   If we
ignore feedback and  the star formation  threshold  density, such that
{\it   all} the  available baryons    are transformed   into a  stable
disk/bulge system,  our models indeed yield  a TFR with $\gamma  = 3$,
but with a large amount of scatter (Figure~1, model~L0).  This scatter
owes to the spread in the angular momenta, J, of proto-galaxies: halos
with  lower J yield more compact  disks  and, because of the adiabatic
contraction,  more   concentrated  halos.   Consequently, less rapidly
spinning proto-galaxies   result  in   disks  with   higher   rotation
velocities.

Taking the  stability related star  formation threshold densities into
account increases  $\gamma$ from $3.0$  to $3.6$ (Figure~1, model~L1). 
In addition, the  scatter is strongly  reduced.  This owes to the fact
that more compact  disks  have  higher  disk mass  fractions that  are
eligible for star formation,  resulting in brighter disks.  The spread
in  J  therefore induces    a   spread along   the  TFR,  rather  than
perpendicular to it.

Additional  physics are  required to  further  steepen the  TFR to its
observed slope  of $\gamma =  4.2$. In van  den Bosch (1999)  we argue
that feedback is the only feasible mechanism  to achieve this. We have
tuned  the model parameters that  control the feedback from supernovae
to   tilt the TFR   to its observed slope.   The  resulting model (L5)
predicts  an  amount of scatter   that is in  excellent agreement with
observations (see panel on the right in Figure~1).

In order  to  assess the  robustness  of the  resulting model, we  now
compare model L5 to other  independent  observations.  In Figure~2  we
plot the gas mass  fractions,  $M_{\rm HI}/L_B$,  as function  of both
absolute magnitude and central  surface brightness.  The models are in
excellent agreement with  the  data. This  success owes mainly  to the
star formation recipe used,  which yields lower  gas mass fractions in
more compact disks, as observed.  The panels on  the right in Figure~2
plot the characteristic mass-to-light  ratio $\Upsilon_0$ (see van den
Bosch   \&  Dalcanton 1999 for   details)  as function  of the central
surface brightness.   Once again, the  model is in good agreement with
the data, nicely   reproducing the observed   $\Upsilon_0$--$\Sigma_0$
``conspiracy'' (cf. McGaugh \& de Blok 1998).

McGaugh (1998) has shown that  mass discrepancies in disk galaxies set
in at  a characteristic acceleration  of $\sim 10^{-10}  \, {\rm m} \,
{\rm  s}^{-2}$.  In Figure~3  we plot the enclosed mass-to-light ratio
of 40  randomly  chosen galaxies  from model  L5  (each sampled  at 15
different radii) as function  of radius, orbital frequency,  and local
acceleration.   As observed, the    model  galaxies  reveal  a  narrow
correlation between mass-to-light  ratio and acceleration.   This is a
remarkable  success for  the dark matter   model;  there is  no obvious
reason   why  disks in dark     halos  would reveal a   characteristic
acceleration, unlike in the case of modified Newtonian dynamics, where
it is integral to the theory.

\begin{figure}
\plotone{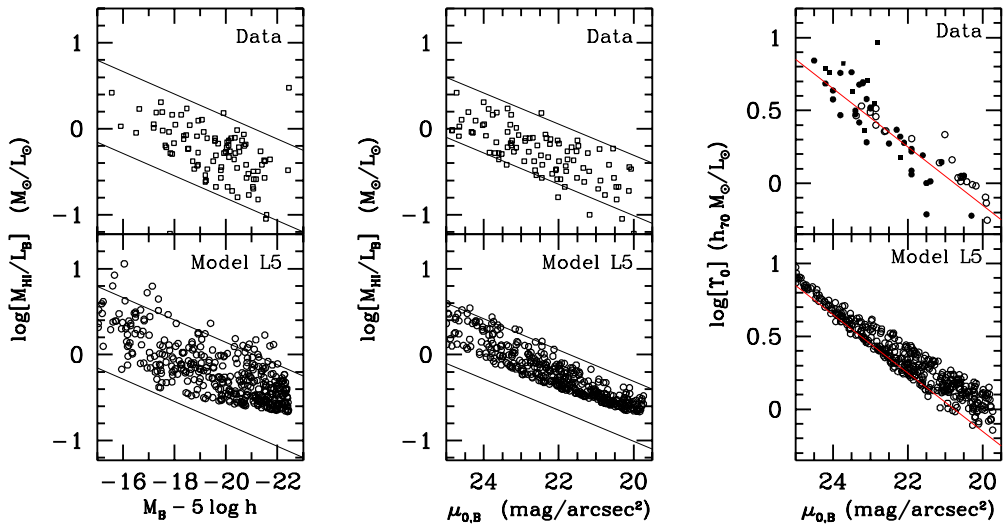}
\caption{Gas mass fractions as function  of absolute magnitude  (left)
  and central  surface brightness  (middle).  The  data is  taken from
  McGaugh \& de Blok (1997), and  is in excellent agreement with model
  L5. Panels on the  right plot the characteristic mass-to-light ratio
  $\Upsilon_0$ as  function of central  surface  brightness.  The data
  (compiled from various sources, see van den Bosch \& Dalcanton 1999)
  reveals   a  narrow    relation   (the  $\Upsilon_0$   -  $\Sigma_0$
  ``conspiracy''),  which is nicely reproduced by  the model. The thin
  lines   have no physical   meaning,  but  are plotted to  facilitate
  comparison between data and model.}
\end{figure}

\begin{figure}
\plotone{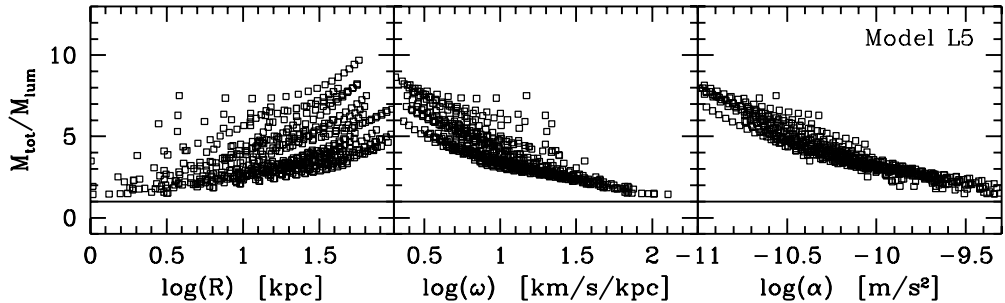}
\caption{The enclosed mass-to-light ratio $M_{\rm tot}/M_{\rm lum}$ as
  function of radius  $R$ (left), orbital  frequency $\omega$ (middle)
  and local acceleration  $\alpha$ (right) for  galaxies of model  L5. 
  The scatter  in $M_{\rm tot}/M_{\rm  lum}$ is minimized when plotted
  versus $\alpha$, indicative of a characteristic acceleration, and in
  good agreement with observations (see McGaugh 1998).}
\end{figure}

\section{Conclusions}

We have shown that simple models for the formation of disk galaxies in
a  dark matter  scenario can explain  a  wide variety of observations.
After tuning the feedback parameters to fit the slope of the empirical
$K$-band TFR, the  model  predicts gas mass fractions,  characteristic
accelerations, an $\Upsilon_0$ - $\Sigma_0$ ``conspiracy'', and global
mass-to-light ratios  that  are   all  in  excellent  agreement   with
observations,  {\it without  additional  tweaking of the  parameters}.
This  strongly contrast with  the  picture  drawn by  McGaugh  (1999).
Although the   results presented here    may appear a  baby step  (cf.
McGaugh 1999) to some  advocates of modified Newtonian dynamics,  they
can be considered a giant leap for those who believe in dark matter.

\acknowledgments

It is  a  pleasure to thank   Julianne  Dalcanton for  a  fruitful and
enjoyable collaboration.  Financial support has  been provided by NASA
through Hubble Fellowship grant \# HF-01102.11-97.A.

\end{document}